# Transfer-learning-based Surrogate Model for Thermal Conductivity of Nanofluids

Saeel S. Pai*, Abhijeet Banthiya
School of Mechanical Engineering, Purdue University
West Lafayette, IN 47906, USA
Email*: pai15@purdue.edu

**Abstract:** Heat transfer characteristics of nanofluids have been extensively studied experimentally, numerically, and theoretically since the 1990s. Research investigations show that the suspended nanoparticles significantly alter the suspension's thermal properties and heat transfer characteristics. The thermal conductivity of nanofluids is one of the properties that has been widely studied and is generally found to be greater than that of the base fluid. This increase in thermal conductivity is found to depend on several parameters like the size and shape of nanoparticles, their concentration, the individual thermal conductivity of the particles and fluids, etc. Several theories and semi-empirical correlations have been proposed to model the thermal conductivities of nanofluids. However, there is no reliable universal theory/correlation yet to model the anomalous thermal conductivity of nanofluids [1]. In recent years, supervised data-driven methods have been successfully employed to create surrogate models across various scientific disciplines, especially for modeling difficult-to-understand phenomena. These supervised learning methods allow the models to capture highly non-linear phenomena. In this work, we have taken advantage of existing correlations and used them concurrently with available experimental results to develop more robust surrogate models for predicting the thermal conductivity of nanofluids. Artificial neural networks are trained using the transfer learning approach to predict the thermal conductivity enhancement of nanofluids with spherical particles for 32 different particle-fluid combinations (8 particles materials – $Al_2O_3$, CuO, SiC, $TiO_2$, $SiO_2$, MgO, ZnO, Fe, and 4 fluids – pure water, pure ethylene glycol (EG), 40-60 mixture of water and EG, and a 60-40 mixture of water and EG). The large amount of lower accuracy data generated from correlations is used to coarse-tune the model parameters, and the limited amount of more trustworthy experimental data is used to fine-tune the model parameters. The transfer learning-based models' results are compared with those from baseline models which are trained only on experimental data using a goodness of fit metric ($R^2$). It is found that the transfer learning models perform better with $R^2$ values of ~0.93 as opposed to ~0.83 from the baseline models.

**List of symbols**

| | |
|---|---|
| k | Thermal conductivity |
| ∅ | Particle volume fraction |
| Re | Reynolds number |
| Pr | Prandtl number |
| d | Diameter |
| $c_p$ | Specific heat capacity |
| ρ | density |
| $k_B$ | Boltzmann constant |
| T | Temperature |
| ρ | Density |
| μ | Dynamic viscosity |
| ν | Kinematic viscosity |
| Ψ | Sphericity |
| λ | Mean free path |

**Subscripts**

| | |
|---|---|
| eff | Nanofluid |
| p | Nanoparticle |
| f | Base fluid |
| cl | Cluster |
| d | Diameter |

# I. Introduction and Motivation

With increasing demand for computing power complemented by the miniaturization of electronic devices, the power densities in microprocessors and other heat-generating components have been on the rise. The focus has shifted from conventional air-cooling to liquid cooling and two-phase cooling solutions to facilitate heat removal from high-power devices. Under liquid cooling, several studies have been carried out to enhance the heat transfer characteristics of the cooling fluid. The thermal conductivity of the fluid is understandably a crucial factor in improving heat transfer performance. High thermal conductive fluids will allow the development of more efficient cooling systems.

It has been found that a suspension of small solid particles in a dielectric cooling fluid can increase the thermal conductivity of the fluid. However, these suspensions only becomes stable when the particle sizes are small enough (typically < 100 nm). Such fluids with the suspension of nanometer-sized solid particles are called nanofluids. The thermal conductivity of nanofluids has been widely studied and is generally found to higher than that of the base fluid. This increase in thermal conductivity depends on several parameters like the shape and size of suspended nanoparticles, their concentrations, the individual thermal conductivities of the particles and fluids, temperature, etc. Several theories and semi-empirical correlations have been developed to model the thermal conductivities of nanofluids [1]. However, there is no reliable universal theory/correlation that models the anomalous thermal conductivity of nanofluids to match experimental observations, which, despite inconsistencies, are considered by the research community to be more reliable than the correlation predictions. On the experimental front, a thermal conductivity enhancement of as high as 160% over base fluid (i.e., thermal conductivity ratio of 2.6 over the base fluid) has been reported in 1 vol% multiwalled carbon nanotubes (MWCNT) in oil nanofluid [2]. Thus, it can be seen why the thermal community remains ever interested in nanofluids research.

To use the enhanced thermal conductivity property of the nanofluid in cooling applications, it is essential to correctly predict the enhancement based on prior knowledge of the parameters involved. In the absence of a complete physical knowledge of the phenomenon, data-driven methods seem like a reasonable choice. In recent years, supervised machine learning (ML) methods have been successfully used to create surrogate models across various scientific disciplines, especially for modeling difficult-to-understand phenomena. These supervised learning methods allow the models to capture highly non-linear phenomena. Recognizing these advantages, researchers have tried using machine learning methods to create surrogate correlations for predicting the thermal conductivity enhancement in nanofluid [3]. However, these works have two glaring gaps. The first being that all currently available surrogate models are trained exclusively on experimental data. However, experimental data are limited because of which generalized models are difficult to develop, which leads us right to the second gap. Most models are developed only for a particular particle-fluid combination (e.g., $Al_2O_3$-$H_2O$). There are very few models available that work for several particle-fluid combinations.

In this work, we try to bridge these gaps by taking advantage of existing correlations/models and using them concurrently with available experimental results to develop more robust surrogate models for predicting the thermal conductivity of nanofluids. There are many correlations available in literature, which, although not very accurate, give decent results, and follow known trends across parametric variation in size, temperature, etc. A large amount of data relating the particle and fluid properties to the thermal conductivity of the nanofluid can be generated by doing a parametric sweep on these correlations. Thus, on the one hand, we have a large amount of data that are not very accurate, while on the other hand, we have a small amount of experimental data that are more accurate. We use the method of transfer learning [4] to effectively use both the low-quality data from correlations and the high-quality data from experiments to develop more robust surrogate models for predicting the thermal conductivity enhancement of several particle-fluid nanofluid combinations. As traditional machine learning approaches are entirely data-driven, the best surrogate models that can be developed are restricted by the quality of data used to train the models. As data are so important, we first review the experimental and theoretical/empirical studies in literature from where we obtain the data before moving on to the development of the surrogate models.

## II. Review of Experimental and Theoretical Studies

### 1. Experimental studies

Many experimental studies have been published in the past few decades, underscoring nanofluids' heat transfer enhancing properties over the base fluid. However, many parameters affect the thermal conductivity of nanofluid [1], [5], sometimes leading to contradictory results from one research group to another, if some of the factors are not accounted for. The experiments study the influence of various physical parameters that affect the thermal conductivity of nanofluids. The main factors that are found to influence the thermal conductivity of nanofluids are 1) particle material, 2) base fluid, 3) volume fraction of the particles in the fluid, 4) temperature, 5) size of the nanoparticle, 6) particle shape, 7) amount of clustering and 8) pH of the fluid. Other factors also affect the stability and, in return, the thermal conductivity of nanofluids such as the surfactant used for stabilizing the solution, the duration of the ultrasonic pulse during the preparation of the nanofluids, etc.; however, not many research papers give these details and are hence not considered in the current surrogate model development. $Al_2O_3$, $CuO$, $TiO_2$, $SiO_2$, $MgO$, and $ZnO$ are some of the commonly encountered nanoparticle materials, while most of the experiments are carried out with distilled water, a mixture of water and ethylene glycol, or engine oil as the base fluid.

The method used for the experiment also played a role in determining the thermal conductivity value. Most studies used a modified transient hot-wire method to measure the thermal conductivity of nanofluids; however, few studies utilize other methods such as steady-state parallel plate techniques, temperature oscillation techniques, etc. Most experiments show a linear increase in thermal conductivity with increasing volume fraction, but this is not always the case. Similarly, many experiments have found that the thermal conductivity of the nanofluid increases with increasing temperature, but again, there are some contradictory results as well. In general, smaller and longer particles were found to enhance the thermal conductivity of the nanofluid more than larger spherical particles. In all these experiments, the thermal conductivity enhancement is expressed as the ratio of the nanofluid's thermal conductivity to the base fluid's thermal conductivity. In the rest of the text, we shall refer to this enhancement in the thermal conductivity as simply the thermal conductivity ratio.

### 2. Theoretical models

Along with the experimental studies, there has also been much theoretical work of increasing complexity done to develop models that can predict the thermal conductivity of nanofluids. Based on the different hypotheses used to explain the enhancement in thermal, these models can be grouped into the following types – old models, Brownian motion models, particle clustering models, and liquid layering models. Some researchers have also explored the effects of ballistic phonon transport and near-field radiation. The old models are the Maxwell [6] and Hamilton and Crosser [7] models, developed initially for milli to micro-scale mixtures of particles and fluid but are still commonly used in literature due to their simplicity. The Brownian motion models like those by Jang and Choi [8] and Koo and Kleinstreuer [9] use the concept of Brownian motion to explain the enhancement in the thermal conductivity of nanofluid. Particle clustering models examine the effect of clustering and resulting interfacial thermal resistance, like the three-step analysis using the Bruggeman model [10] and the model by Nan et al. [11]. Other models that follow this approach are developed by Xuan et al. [12] and Chen et al. [13]. Liquid layering models like those by Yu and Choi [14], Xie et al. [15], and Xue and Xu [16] study and model the influence of the formation of a liquid layer around the nanoparticles on the thermal conductivity of the nanofluid. While none of the models gave a perfect match with the experimental results, it was seen from the literature [1], [5] that the models by Jang and Choi, Koo and Kleinstreuer, Xie et al., and Yu and Choi were found to have a good agreement with at least some subsections of experimental data. Moreover, Koo and Kleinstreuer's and Jang and Choi's models could also predict the increase in the thermal conductivity ratio with increasing temperature, which was seen from many experiments. With this in mind, we have used four models to generate data for training the machine learning model discussed below.

2.1. Hamilton-Crosser model (1962) [7]

Hamilton-Crosser model was initially developed to determine the thermal conductivity of the micro-sized particle and fluid mixture. The model builds upon the Maxwell model, which was developed over a century ago to include particle shape.

Although this model underpredicts the thermal conductivity enhancement, it has been widely used in the literature due to its simplicity and no empirical functionality embedded into the correlation. Hamilton-Crosser model depends on the particle thermal conductivity, base fluid thermal conductivity, and a shape factor to predict the effective thermal conductivity of nanofluid.

$$k_{eff} = \frac{k_p + (n-1) + (n-1)(k_p - k_f)\emptyset}{k_p + (n-1)k_f - (k_p - k_f)\emptyset} k_f$$

where, $n = \frac{3}{\psi}$, $\psi$ is the sphericity of the particle defined as the ratio of the surface area of a sphere with a volume equal to that of the particle to the surface area of the particle. For spheres n=3, the Hamilton-Crosser model becomes identical to Maxwell's thermal conductivity model.

### 2.2. Jang-Choi (2004) [8]

Jang and Choi considered the effect of the Brownian motion of nanoparticles on the effective thermal conductivity of the nanofluid. The model assumes that the heat transfer in the nanofluids occurs through four different modes, which are a) collisions of fluid molecules (conduction in fluid); b) conduction in solid nanoparticle material; c) collisions of nanoparticles; d) micro-convection set up in the base fluid due to Brownian motion of nanoparticles leading to local enhancements in heat transfer. Out of these four modes of heat transfer, it was concluded that the heat transfer through collisions of the nanoparticles was negligible when compared to the other three modes because of a statistically lower number of collisions when compared to fluid-fluid or fluid-nanoparticle interaction. Hence particle-particle heat transfer was neglected in the final correlation.

As this model considers the effect of Brownian motion, the thermal conductivity of nanofluid depends on temperature and particle size. The trends of increasing thermal conductivity with decreasing nanoparticle size and increasing thermal conductivity with increasing temperature followed the trends from experimental results. This is because smaller particles have a higher Brownian motion and higher energy at higher temperatures.

$$k_{eff} = k_f(1 - \emptyset) + k_p^* \emptyset + 3C_1 \left(\frac{d_f}{d_p}\right) k_f Re_d^2 Pr_f \emptyset$$

where $C_1$ is a proportionality constant and $k_p^*$ is the effective particle thermal conductivity after considering Kapitza resistance defined as

$$k_p^* = \beta k_p$$

where $\beta \sim O(10^{-2})$ is a constant. Reynolds number is defined using random motion velocity $\bar{C}_{R.M.}$ kinematic viscosity of the base fluid.

$$Re_d = \frac{\bar{C}_{R.M.} d_p}{\nu_f}$$

$$\bar{C}_{R.M.} = \frac{D_0}{\lambda_f}$$

$\lambda_f$ is the fluid molecule mean free path and $D_0$ is nanoparticle diffusion coefficient defined as:

$$D_0 = \frac{k_B T}{3\pi\mu_f d_p}$$

Note that the proportionality constant C[1] and β were not provided by the group and were fine-tuned by us to fit the experimental data available for different particle-fluid combinations discussed in section III.

2.3. Koo and Kleinstreuer (2004) [9]

This model assumes that the net thermal conductivity enhancement can be divided into two components, one from static enhancement (obtained using classical Maxwell model[6]) and enhancements due to Brownian motion

$$k_{eff} = k_{static} + k_{Brownian}$$

$$k_{Brownian} = 5 \times 10^4 \beta\phi\rho_f c_{p,f} \sqrt{\frac{k_B T}{\rho_p d_p}} f$$

This model is different from Jang-Choi in the way that the β term is empirically introduced after the derivation of the equation to consider the effect of nanoparticle-base fluid interactions. It was shown that the β term is more effective at higher volume fractions ($\phi$). Also, the parameter f was introduced to increase temperature dependence. Koo and Kleinstreuer gave the functional forms for β and f for certain particle fluid combinations, which are tabulated in their paper, obtained using fitting of experimental data.

2.4. Xie et al. (2005) [15]

Xie et al. modeled the thermal conductivity of nanofluid by considering the effect of a liquid nanolayer around the nanoparticle. The nanolayer is assumed as a hard spherical shell of thickness t around the nanoparticle. The thermal conductivity of this nanolayer shell is assumed to vary linearly from solid thermal conductivity close to the nanoparticle surface and base fluid conductivity at the nanolayer-fluid interface.

$$\frac{k_{eff} - k_f}{k_f} = 3\Theta\phi_T + \frac{3\Theta^2 \phi_T^2}{1 - \Theta\phi_T}$$

$$\Theta = \frac{\beta_{lf}\left[(1+\gamma)^3 - \beta_{pl}/\beta_{fl}\right]}{(1+\gamma)^3 + 2\beta_{lf}\beta_{pl}}$$

Here, $\beta_{lf}$, $\beta_{pl}$, and $\beta_{fl}$ are parameters dependent on thermal conductivity of particle, base fluid, and nanolayer given as

$$\beta_{lf} = \frac{k_l - k_f}{k_l + 2k_f}$$

$$\beta_{pl} = \frac{k_p - k_l}{k_p + 2k_l}$$

$$\beta_{fl} = \frac{k_f - k_l}{k_f + 2k_l}$$

And $k_l$ was defined as:

$$k_l = \frac{k_f M^2}{(M - \gamma)ln(1 + M) + \gamma M}$$

$$M = \frac{k_p(1 + \gamma)}{k_f} - 1$$

$\phi_T$ is the updated total volume fraction of nanoparticles and nanolayer

$$\phi_T = \phi(1 + \gamma)^3$$

$$\gamma = \frac{t}{r_P}, \quad \text{where } t \text{ is thickness of nanolayer}$$

Different thicknesses of nanolayers gave different values for thermal conductivity. Hence, the thickness of the nanolayer was assumed to be a model parameter that was fitted for using experimental results.

## III. Data Collection

From the experimental works and the theoretical models that were considered, particle material, base fluid, volume fraction of the particles in the fluid, temperature, and size of the nanoparticles were found to be the five most encountered parameters. Thus, we use these five parameters as the inputs to the machine learning model, whereas the output is the thermal conductivity ratio of the nanofluid. In this work, the shape of the particle has not been taken into consideration, assuming the particles to be spherical.

### 1. Experimental data

Experimental data were extracted from nineteen different highly cited papers. Table 1 lists out the papers from which the experimental data was collected, along with the parametric variations considered in those papers. 1015 distinct data points on eight different particle materials and four different base fluids have been collected. The eight particle materials were $Al_2O_3$, CuO, SiC, $TiO_2$, $SiO_2$, MgO, ZnO, and Fe, and the four fluids were pure water, pure ethylene glycol (EG), 40-60 mixture of water and EG, and a 60-40 mixture of water and EG. The volume concentration limit was typically between 0-10% (although some data points have concentrations going up to 20%), particle sizes are less than 100 nm, and the temperatures typically vary between 20-60 °C.

**Table 1. List of sources of experimental data and the parametric variations considered**

| No. | Year | Authors | Particles Material | Base Fluid | Volume Conc. [%] | Sizes [nm] | Temp [°C] |
|---|---|---|---|---|---|---|---|
| 1 | 1999 | Lee et al. [17] | $Al_2O_3$, CuO | water, EG | 0-4 | 38.4, 23.6 | 25 |
| 2 | 1999 | Wang et al. [18] | $Al_2O_3$, CuO | water, EG | 0-8 | 28, 23 | 24 |
| 3 | 2002 | Xie et al. [19] | SiC | water, EG | 0-5 | 26 | 4 |
| 4 | 2004 | Murshed et al. [20] | $TiO_2$ | water | 0-5 | 15 | 25 |
| 5 | 2006 | Li, Peterson [21] | $Al_2O_3$, CuO | water | 0-10 | 29, 36 | 27-35 |
| 6 | 2006 | Kang et al. [22] | $SiO_2$ | water | 0-5 | 20 | 21-23 |
| 7 | 2008 | Mintsa et al. [23] | $Al_2O_3$, CuO | water | 0-20 | 29, 36, 47 | 20-50 |
| 8 | 2009 | Vajjha, Das [24] | $Al_2O_3$, CuO, ZnO | 60:40 EG/W | 0-8 | 53, 29, 77 | 20-95 |
| 9 | 2009 | Duangthongsuk, Wongwises [25] | $TiO_2$ | water | 0-2 | 21 | 15-35 |
| 10 | 2010 | Timofeeva et al. [26] | SiC | water | 4.1 | 16-90 | 22.5 |
| 11 | 2010 | Jahanshahi et al. [27] | $SiO_2$ | water | 0-4 | 12 | 25 |
| 12 | 2010 | Moosavi et al. [28] | ZnO | EG | 0-3 | 70 | 10-50 |
| 13 | 2013 | Hussein et al. [29] | $Al_2O_3$, $TiO_2$, $SiO_2$ | water | 1-2.5 | 13, 30, 30 | 25-50 |
| 14 | 2013 | Reddy, Rao [30] | $TiO_2$ | water, 40:60 EG/W | 0-1 | 21 | 30-70 |
| 15 | 2013 | Barbes et al. [31] | $Al_2O_3$ | water, EG | 0-10 | 45 | 25-65 |
| 16 | 2014 | Esfe et al. [32] | MgO | EG | 0-5 | 20-60 | 25-55 |
| 17 | 2015 | Esfe et al. [33] | Fe | water | 0-1 | 37, 71, 98 | 25 |
| 18 | 2015 | Esfe et al. [34] | MgO | water, 40:60 EG/W | 0-3 | 40 | 20-50 |
| 19 | 2015 | Esfe et al. [35] | $Al_2O_3$ | EG | 0-5 | 5 | 24-50 |

## 2. Data from models

By varying the five parameters discussed above (particle material, base fluid, volume fraction of the particles in the fluid, temperature, and size of the nanoparticles), ~45000 distinct data points are generated using the models discussed in Section II. Note that some models are valid only for a few particle/base fluid systems as they require empirical function fittings based on experimental data, which are available only for limited particle/fluid combinations in the literature. Also, the models might be valid only for a small range of parameters. Such models have been used with precaution, keeping all those model limitations in mind to generate the training data.

**Table 2. List of models used and the parametric variations considered**

| No. | Authors | Particles Material | Base Fluid | Volume Conc. [%] | Sizes [nm] | Temp [°C] | Data Points |
|---|---|---|---|---|---|---|---|
| 1 | Hamilton-Crosser | $Al_2O_3$, CuO, Fe, $TiO_2$, MgO, SiC, $SiO_2$, ZnO | water, EG, 40:60 EG/W, 60:40 EG/W | 0.5-6 | Independent | 20-60 | 3456 |
| 2 | Jang & Choi | $Al_2O_3$, CuO | water, EG | 0.5-5 | 10-150 | 20-60 | 3196 |
| 3 | Koo & Kleinstreuer | $Al_2O_3$, CuO | water, EG, 40:60 EG/W, 60:40 EG/W | 1-5 | 10-150 | 30-50 | 3887 |
| 4 | Xie et al. | $Al_2O_3$, CuO, Fe, $TiO_2$, MgO, SiC, $SiO_2$, ZnO | water, EG, 40:60 EG/W, 60:40 EG/W | 0.5-5 | 10-150 | 20-60 | 34560 |

Hamilton-Crosser model is independent of particle size and requires only the thermal conductivities of particle and base fluid as the inputs. For completeness of the training data however, for each particle material, the average particle size as observed from the experimental data is used. Although this model is independent of temperature, the thermal conductivities of the base fluid are varied with temperature. All eight particles and four fluid combinations were used in Hamilton-Crosser and Xie et al. models. For Jang & Choi's model, only three-particle/fluid combinations were used, $Al_2O_3$-$H_2O$, CuO-$H_2O$, CuO-EG. 4 Fluid, and two particles (total 8) combinations were used in Koo and Kleinstreuer's Model as this model requires empirical parameters (β, f) which are a function of particle type. To generate the data, the sizes have been varied between 10-150 nm, but experimental data, which is used to fine tune the model parameters, only has sizes between 15-100 nm. Furthermore, as will be seen later, the performance of the models developed has been tested on experimental data. Thus, the models developed are claimed to be valid only for spherical particles in the size range of 15-100 nm.

## IV. Surrogate Model Development

Transfer learning [4] is a category of machine learning algorithms that focuses on storing knowledge gained while solving one problem and applying it to a different but related problem. Here, an ML method developed for one task using one set of data is used for another similar task by modifying the already trained model weights using new data relevant to the second task. Transfer learning is more commonly used in computer vision where pre-trained networks like the Google Inception Model or Microsoft ResNet Model, trained on millions of images, are used for a specific application by adapting them accordingly. For a feedforward neural network, this adaptation is typically made by taking a fully trained neural network, freezing the weights in all its layers except in the last one or few layers, and then training the weights in the unfrozen last few layers on the new task-specific data. The understanding here is that the first few layers of an artificial neural network or a convolutional neural network help identify the data's overall trends and broad features, whereas the last few layers learn the specific details relevant to the tasks at hand. The advantage of this is that now only a few data samples are needed to fine-tune the models (i.e., train the last few layers) to suit a specific application that is in some way similar to the one the original model is developed for (e.g., computer vision). The bulk of the training is done on a large volume of images that are not closely related to the specific task but help the model learn the basics of image classification.

Our application has a large volume of low-accuracy data from the models and correlations, and a smaller amount of more accurate (considered to be) experimental data. Using similar principles as explained above, we can use the large volume of low-quality data to coarse tune our ML model to learn the trends in the thermal conductivity enhancement of nanofluids, and then use the small quantity of experimental data to fine-tune the model to improve the prediction accuracy for our task. An artificial neural network (ANN) has been used as the ML model in this work. The structure of an ANN used in the work is shown in figure 1. An ANN is a directed acyclic graph. The leftmost layer of the ANN (in figure 1) is called the input layer, the rightmost layer is called the output layer, and the layers in between are the hidden layers. In our case, the ANN takes in 5 inputs, which are 1) particle material, 2) base fluid, 3) volume fraction of the particles in the fluid, 4) temperature, 5) size of the nanoparticle. The model is trained on eight different particle materials - $Al_2O_3$, $CuO$, $SiC$, $TiO_2$, $SiO_2$, $MgO$, $ZnO$, and $Fe$, and four different base fluids - pure water, pure ethylene glycol (EG), 40-60 mixture of water and EG, and a 60-40 mixture of water and EG. Particle material and base fluid, which are categorical data, have been encoded using one hot vectors. Thus, the one-hot vectors corresponding to particle material and the base fluid are eight-dimensional and four-dimensional vectors respectively. The volume fraction, temperature and the size of nanoparticles are numerical data which have been normalized for use in the ANN. The input, therefore, is ultimately a fifteen-dimensional vector. The output of the ANN is the thermal conductivity ratio, which is a scaler. The ANN is figure 1 has 3 hidden layers. To see how well the transfer learning approach works, the results from the transfer learning models have been compared to the results obtained from models trained only on experimental data (called the baseline models). Furthermore, the best performing transfer learning models are aggregated to form an ensemble model, which helps is reducing prediction uncertainty.

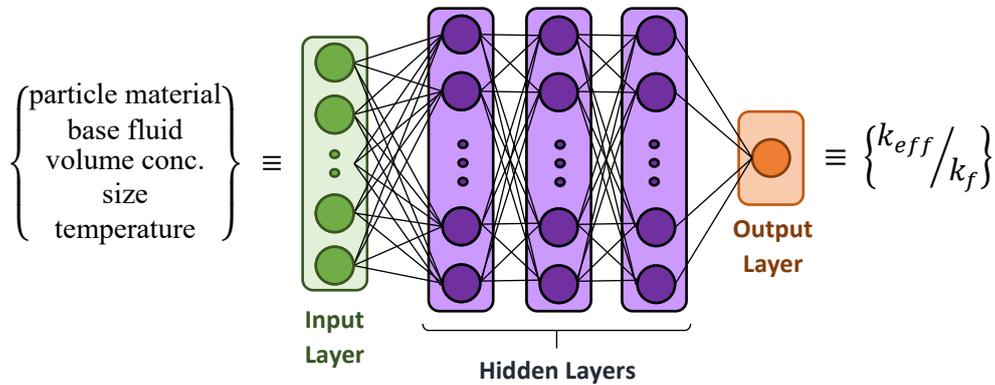

**Figure 1.** Structure of an artificial neural network (ANN). The input is a fifteen-dimensional vector, and the output is the ratio of thermal conductivities of the nanofluid and the base fluid

1. Training and Hyperparameter Optimization

The dataset consists of two types of data – 1) data generated from models and correlations (45,099 data points), and 2) data from experiments (1015 data points). The data from correlations is split into two parts, the training and validation data, with 70% of the data forming the training and the remaining 30% forming the validation set. The experimental data, however, is divided into 3 parts, for training, validation and testing in the ration 60:20:20. The baseline models are trained on the experimental training data while their performance during training is monitored by calculating the mean squared error (MSE) on the experimental validation data. The final performance is evaluated on the experimental test data which is completely hidden from the model during the training phase. The transfer learning model, on the other hand, is first trained on the training data from correlations while the performance is monitored on the validation data from correlations. Next, all the weights except the weights corresponding to the last layer are frozen and the unfrozen weights are trained again using the experimental training data and performance monitored on the experimental validation data. The final performance of the transfer learning model is also evaluated on the experimental test data, as in the case of the baseline mode. The reason for this is after all, we want the model to be precise on the experimental data and not necessarily on the data generated from correlations.

The baseline and the transfer learning models are developed in Python using TensorFlow and Keras libraries. ReLU has been used as the activation function in the hidden layers. MSE has been used as the loss function with Adam (learning rate of $1 \times 10^{-3}$) as the weight optimization algorithm. Batch normalization was used each hidden layer to reduce the training time, whereas a dropout rate of 0.25 was used after each hidden layer to reduce overfitting. A batch size of 64 was used during the optimization process. The hyperparameters mentioned above were selected based on experimental data. The reason for using experimental data is twofold: 1) we want the model to be precise on the experimental data and not necessarily on the data from correlations, and 2) it is assumed that both the experimental data and the data from correlations follow similar distributions, and hence, the hyperparameters that are optimal for one set would also be optimal for the other, and training models on experimental data is much faster than on data from correlations (~1,000 data points as compared to ~45,000).

ANNs with two and three hidden layers were considered for both, the baseline models and the transfer learning models. The number of nodes in each of the hidden layers was optimized by doing a grid search, with the number of nodes in each layer varying from 10 to 300, and comparing the goodness of fit ($R^2$) value of each model's predictions on the experimental test data. The experimental test set (20% of the total experimental data) was kept the same throughout. The train and validation sets were shuffled between reruns of each model (i.e specified number of hidden layers and number of nodes in each hidden layer), and the rerun of the model giving the best performance was saved. The training of correlation and experimental data was found to converge within 100 and 500 epochs respectively. Hence these were the number of epochs used in training all the models.

## V.   Results and Discussion

Using the data splits and hyperparameters mentioned in the previous section, the baseline and transfer learning models were trained. The architectures and performances of the top five performing baseline models and transfer leaning models are summarized in tables 3 and 4 respectively. The tables show the number of hidden layers in the ANN, number of nodes in each of those hidden layers, and the $R^2$ values of the models on only the experimental test data, as well as on the entire experimental data (train + validation + test). The last row of both the tables shows the same statistics for an ensemble model that is made by combining the five top performing models. The best performing individual baseline models have $R^2$ values of around 0.83 on both, the experimental test set as well on the entire experimental data set, whereas the baseline ensemble model has slightly higher $R^2$ values of 0.85 and 0.83 respectively on those two data sets. In comparison, the best performing individual transfer learning models have $R^2$ values of around 0.92 on the two data sets. The transfer learning ensemble models in this case do not show a significant increase in $R^2$ values as compared to the individual models. Thus, the transfer learning models clearly outperform the baseline models.

Table 3. Architectures and performance summary of baseline models

| No | # hidden layers | # nodes in hidden layers | $R^2$ value on experimental test data | $R^2$ value on full experimental data |
| --- | --- | --- | --- | --- |
| 1 | 2 | 250-100 | 0.8336 | 0.8174 |
| 2 | 2 | 250-250 | 0.8351 | 0.8299 |
| 3 | 2 | 200-50 | 0.8246 | 0.7919 |
| 4 | 3 | 250-250-150 | 0.8042 | 0.8112 |
| 5 | 3 | 250-150-150 | 0.8040 | 0.7887 |
| 6 | Ensemble of previous 5 models | | 0.8513 | 0.8338 |

Table 4. Architectures and performance summary of transfer learning models

| No | # hidden layers | # nodes in hidden layers | $R^2$ value on experimental test data | $R^2$ value on full experimental data |
| --- | --- | --- | --- | --- |
| 1 | 2 | 300-300 | 0.9284 | 0.9202 |
| 2 | 3 | 150-150-150 | 0.9187 | 0.9103 |
| 3 | 3 | 250-250-150 | 0.9173 | 0.9264 |
| 4 | 2 | 250-250 | 0.9149 | 0.9191 |
| 5 | 2 | 250-150 | 0.9115 | 0.9151 |
| 6 | Ensemble of previous 5 models | | 0.9274 | 0.9285 |

Figures 2 shows the correlation plots for the best performing individual and ensemble models, on the experimental test data. The plots in red and cyan correspond to the baseline and transfer learning-based models respectively. The first and second rows of figure 2 show the scatter plots of the best performing individual and ensemble models respectively. For each of the plots, the x-axis shows the thermal conductivity enhancement measured from true experiments, and the y-axis shows the thermal conductivity enhancement predicted by the ML models on the same data. For a perfect correlation, all the scatter points should lie exactly on the 45-degree line. The dotted lines are set at an offset of 0.05. It is seen that the transfer learning-based models give much better correlation plots – the spread in the scatter plot is lesser and most of the points lie in the space between the dotted lines. Figure 3 shows similar correlation plots for the baseline and transfer learning ensemble models, but on the entire experimental data as opposed to only on the experimental test set. Here too, it is seen that the transfer learning-based model outperforms the baseline model by a large margin. Thus, the transfer learning-based models are clearly the superior models.

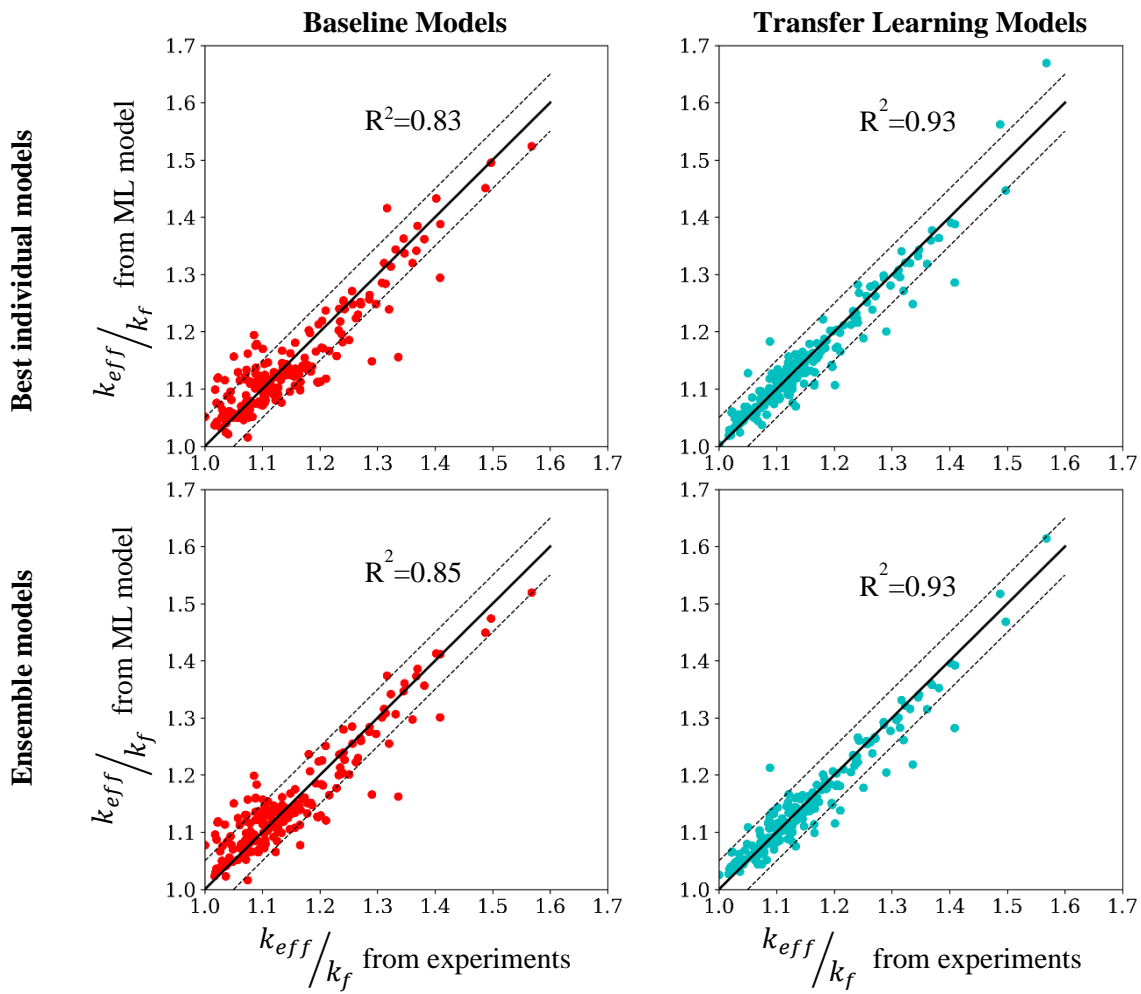

**Figure 2.** Performance of the ML models on the test partition of the experimental data. (Left column): performance of the baseline models (best individual and ensemble) on experimental test data, and (right column): performance of the transfer learning models (best individual and ensemble) on experimental test data

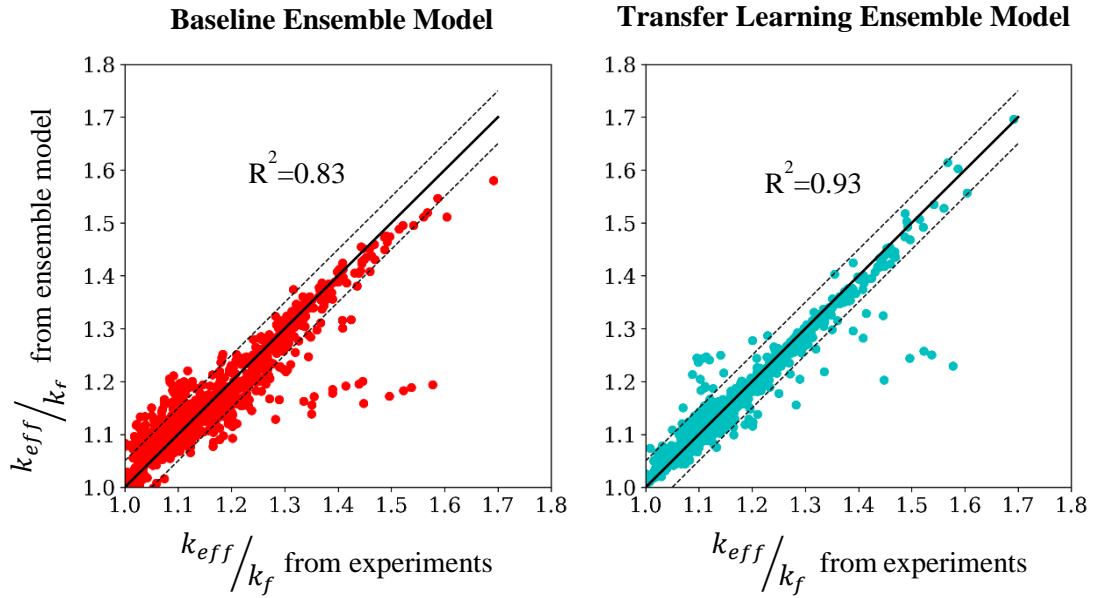

**Figure 3.** Performance of the baseline and transfer learning-based ensemble models on the entire experimental data set. The plot on the left is for the baseline ensemble model and the plot on the right is for the transfer learning-based ensemble model.

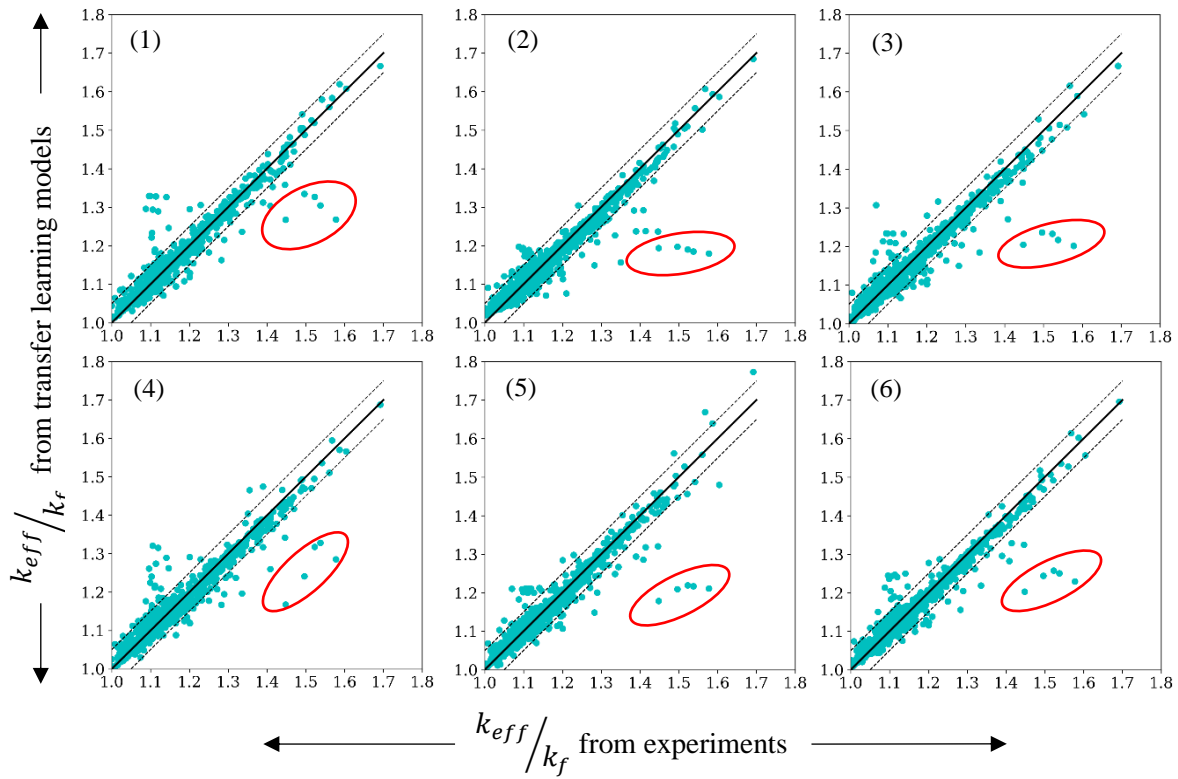

**Figure 4.** Correlation plots for the five top performing transfer learning-based models and the transfer learning-based ensemble model on the entire experimental data. The plots are numbered as per the list in table 4. The plots marked in a red ellipse are the outliers noted in every model's predictions.

Figure 4 shows the correlation plots for the five top performing transfer learning models as well as for the transfer learning-based ensemble model on the entire experimental dataset. These are still the models that are trained on the training section of the experimental dataset (60% of the data). The plots are numbered as per the list in table 4. Like in the previous figures, for each of the plots, the x-axis shows the thermal conductivity enhancement measured from actual experiments, and the y-axis shows the thermal conductivity enhancement predicted by the transfer learning-based models on the same data. It is seen that there are a few points on the right of the 45-degree line, marked in red, that appear as outliers in all 6 plots. These are points that are mispredicted in every model. These data points do not follow the trends that are shown by most other data points. One of the explanations for this could be that the experimental data for these experiments are not accurate. It was mentioned earlier in the paper that sometimes there are discrepancies between experimental results due to some experimental factors not being considered. Although this has not been verified, it could be that this could be one such case. Thus, well trained ML models may also help in identification of erroneous training data. All the compiled data and ML models used in this work are available in the GitHub repository[1].

## VI. Conclusion

In this study, ANNs are trained to predict the thermal conductivity enhancement in nanofluids based on the transfer learning approach. The large amount of low-accuracy data from existing correlations is used to coarse tune the weights of the ANN, whereas the small amount of more accurate experimental data is used to fine tune the weights of the ANN. The ML models are developed for spherical nanoparticles with sizes less than 100 nm and are trained on eight different nanoparticle materials ($Al_2O_3$, CuO, SiC, $TiO_2$, $SiO_2$, MgO, ZnO, and Fe) and four base fluids (pure water, pure EG, 40-60 mixture of water and EG, and a 60-40 mixture of water and EG). Thus, the ML models are valid for a total of 32 particle-fluid combinations. The volume concentration of the nanoparticles was typically kept between 0-10% and the temperatures were varied between 20-60 °C. The correlation data was generated from four well known correlations, and the experimental data was collected from nineteen highly cited experimental papers. The predictions of the individual and ensemble transfer learning-based models were compared with baseline individual and ensemble models which were trained only on experimental data using a goodness of fit metric. All the transfer learning-based models were found to perform much better than the baseline models, with $R^2$ values of

---

[1] GitHub repository: https://github.com/SaeelPai/nanofluids-therm-cond-ml-models

~0.93 as opposed to ~0.83 from the baseline models. The transfer learning-based models were found to be more robust than models trained on only experimental data. Apart from this, there is also the possibility of using well trained ML models to identify potentially erroneous training data. Thus, the transfer learning approach for developing surrogate models for predicting the thermal conductivity of nanofluids holds great promise and should be explored further.